# Simulating non-small cell lung cancer with a multiscale agent-based model

Zhihui Wang, Le Zhang, Jonathan Sagotsky, and Thomas S. Deisboeck [§]

Complex Biosystems Modeling Laboratory, Harvard-MIT (HST) Athinoula A. Martinos Center for Biomedical Imaging, Massachusetts General Hospital, Charlestown, MA 02129, USA.

[§]**Corresponding Author:**

Thomas S. Deisboeck, M.D.
Complex Biosystems Modeling Laboratory
Harvard-MIT (HST) Athinoula A. Martinos Center for Biomedical Imaging
Massachusetts General Hospital-East, 2301
Bldg. 149, 13th Street
Charlestown, MA 02129
Tel: 617-724-1845
Fax: 617-726-7422
Email: deisboec@helix.mgh.harvard.edu

**Email addresses:**

    ZW:    billwang@nmr.mgh.harvard.edu
    LZ:    adamzhan@nmr.mgh.harvard.edu
    JS:    sagotsky@nmr.mgh.harvard.edu
    TSD:    deisboec@helix.mgh.harvard.edu

# Abstract

## Background

The epidermal growth factor receptor (EGFR) is frequently overexpressed in many cancers, including non-small cell lung cancer (NSCLC). *In silcio* modeling is considered to be an increasingly promising tool to add useful insights into the dynamics of the EGFR signal transduction pathway. However, most of the previous modeling work focused on the molecular or the cellular level *only*, neglecting the crucial feedback between these scales as well as the interaction with the heterogeneous biochemical microenvironment.

## Results

We developed a multiscale model for investigating expansion dynamics of NSCLC within a two-dimensional *in silico* microenvironment. At the molecular level, a specific EGFR-ERK intracellular signal transduction pathway was implemented. Dynamical alterations of these molecules were used to trigger phenotypic changes at the cellular level. Examining the relationship between extrinsic ligand concentrations, intrinsic molecular profiles and microscopic patterns, the results confirmed that increasing the amount of available growth factor leads to a spatially *more* aggressive cancer system. Moreover, for the cell closest to nutrient abundance, a *phase-transition* emerges where a minimal increase in extrinsic ligand *abolishes* the proliferative phenotype altogether.

## Conclusions

Our *in silico* results indicate that, in NSCLC, in the presence of a strong extrinsic chemotactic stimulus, and depending on the cell's location, downstream EGFR-ERK signaling may be processed *more* efficiently, thereby yielding a *migration-dominant* cell phenotype and overall, an *accelerated* spatio-temporal expansion rate.

## Background

Non-small cell lung cancer (NSCLC) remains at the top of the list of cancer-related deaths in the United States [1]. The epidermal growth factor receptor (EGFR) is frequently overexpressed in NSCLC [2, 3]. Binding of epidermal growth factor (EGF) or transforming growth factor alpha (TGFα) to the extracellular domain of EGFR produces a number of downstream effects that affect phenotypic cell behavior including proliferation, invasion, metastasis, and inhibition of apoptosis [4]. In particular, increasing the expression of these growth factors leads to EGFR hyperactivity [5, 6], and thus increases tumor cell motility and invasiveness, and finally enhances lung metastasis [7, 8]. Since approximately 90% of all cancer deaths originate from the spread of primary tumor cells into the surrounding tissue [9], quantitative measurements of the relationship between the level of the growth factors and the resulting tumor expansion is crucial - all the more so, since EGFR has emerged as an attractive therapeutic target for patients with advanced NSCLC [10].

A number of EGFR-related intracellular signal transduction pathways have been studied [11-16], including NSCLC [17], and corresponding computational models at the molecular-level have been developed. These quantitative works mainly focused on signal-response relationships between the binding of EGF to EGFR and the activation of downstream proteins in the signaling cascade. With these *in silico* approaches, experimentally testable hypotheses can be made on signaling events controlling divergent cellular responses such as cell proliferation, differentiation, or apoptosis [18, 19]. However, most signaling works did not yet consider the cellular level (see [20, 21] for a review), and, conversely, only a few recent EGF/EGFR-mediated cellular-level models have started to incorporate a simple molecular level in studying e.g., cell migration in breast cancer [22], cell proliferation [23], and autocrine receptor-ligand dynamics [24, 25]. We argue that a more detailed understanding of a complex cancer system requires integrating *both* molecular- and cellular-level works to properly examine multicellular dynamics. To our knowledge, to date, no multiscale model of NSCLC has been developed or published.

Our group has been developing multiscale models to investigate highly malignant brain tumors as *complex dynamic and self-organizing biosystems*. Since this NSCLC model builds on these works, we will briefly review some milestones. First, an agent-based model for studying the spatio-temporal expansion of virtual glioma cells in a two-dimensional (2D) environment was built and the relationship between rapid growth and extensive tissue infiltration was investigated [26, 27]. This 'micro-macro' framework was then extended 'top-down' by incorporating an EGFR molecular interaction network [28] so that molecular dynamics at the protein level could be related to multi-cellular tumor growth patterns [29]. Most recently, an explicit cell cycle description was implemented to study in more detail tumor growth dynamics in a three-dimensional (3D) context of a virtual brain tumor [30]. These previous works have provided a computational paradigm in which biological processes have been successfully simulated from the molecular scale up to the cellular level and beyond. This progress led us to test the platform's applicability to and flexibility for other cancer types as well.

In this paper, we have therefore extended these previous modeling works to the case of NSCLC. Necessary modifications include at the molecular level the implementation of a NSCLC-specific EGFR-ERK signal transduction pathway. A novel, data-driven switch that is operated by two key molecules, i.e. phospholipase Cγ (PLCγ) and extracellular signal-regulated kinase (ERK), processes the phenotypic decision at the cellular level. The aim of this *in silico* work is to provide insights into the externally triggered molecular-level dynamics that govern phenotypic changes and thus impact multicellular patterns in NSCLC. In the following sections, we will first show the detailed design of the model before we present and then discuss the simulation results.

## Model

### Molecular Signaling Pathway

The kinetic model of the implemented NSCLC-specific molecular signaling pathway, which consists of 20 molecules, is shown in **Fig. 1**. These proteins, including both receptor (EGFR) and non-receptor kinases (e.g., PLCγ and protein kinase C (PKC) [31, 32], Raf, mitogen-activated protein kinase kinase (MEK), and ERK [33-35]), have been experimentally or clinically proven to play an important role in NSCLC tumorigenesis. Although in reality these molecules fulfil their functions by interacting with a multitude of other molecular species from many distinct pathways [36, 37], we choose to start with these proteins not only because of their significance in the case of NSCLC but also since most of their kinetic parameters can be found in the literature. Also, it is reasonable to reduce the number of involved molecules as a starting point for modeling [38]. Amongst these proteins, both PLCγ and ERK are of particular interest for determining the cell's phenotypic changes as we will detail below.

**Figure 1**

Kinetic equations are written in terms of concentrations and the reaction rates are functions of concentrations. The association and dissociation steps are characterized by first-order and second-order rate constants, respectively. We note that, although in reality chemical reactions of second or higher order are two-step processes, they are usually treated as a one-step process in mathematical modeling [39]. Our model is based on a total of 20 ordinary differential equations (ODEs) and uses exactly the same modeling techniques as other pathway analysis studies (see [11, 12] for detailed definitions). For simplicity, the ODEs for different molecules were calculated by Eq. (1):

$$\frac{d(X_i)}{dt} = \sum v_{\text{Production}} - \sum v_{\text{Consumption}} \tag{1}$$

where $X_i$ represents one of these 20 molecular pathway components. In Eq. (1), the change in concentration of molecule $X_i$ is the result of the reaction rates producing $X_i$ minus the reaction rates consuming it. Each biochemical reaction is then characterized by $v_i$ (see **Fig. 1**) with forward and reverse rate constants. **Tables 1** and **2** summarize the kinetic parameters and the ODEs used for the model.

**Table 1**

**Table 2**

**Micro-Environment**

The 2D virtual micro-environment is made up of a discrete lattice consisting of a grid with 200 x 200 points (**Fig. 2**). We use *p(i,j)* to express each point in the lattice, where *i* and *j* indicate the integer location in Euclidean terms. One single, distant nutrient source (simulating a cross-sectional blood vessel) is located at *p(150, 150)*. To start with, a number of M x N cells (in other words, an M-by-N matrix) are initialized in the center of the lattice (and this number can be set to meet different simulation purposes). Each grid point can be occupied with one cell only or remain empty at a time.

**Figure 2**

Three external chemical cues are employed in the model: EGF, glucose and oxygen tension. As we have done in previous studies [29, 30], the nutrient source carries the highest value of these three diffusive cues, which implicates that it is the most attractive location for the chemotactically acting tumor cells. Then, by means of normal distribution, each grid point of the lattice is assigned a concentration profile of these three cues. The levels of these distributions are weighted by the distance, $d_{ij}$, of a given grid point from the nutrient source. The distributions of these three cues are described by the following equations:

$$EGF^{ij} = T_m \cdot \exp(-2d_{ij}^2 / \sigma_t^2) \tag{2}$$

$$Glucose^{ij} = G_a + (G_m - G_a) \cdot \exp(-2d_{ij}^2 / \sigma_g^2) \tag{3}$$

$$Oxygen^{ij} = O_a + (O_m - O_a) \cdot \exp(-2d_{ij}^2 / \sigma_o^2) \tag{4}$$

Moreover, the three chemotactic cues continue to diffuse over the lattice throughout the entire process of a simulation with a fixed rate, using the following equation:

$$\frac{\partial M^{ij}}{\partial t} = D_M \cdot \nabla^2 M^{ij}, \quad t = 1,2,3,\ldots . \tag{5}$$

where $M$ represents one of the three external cues, and $t$ represents a time step. The coefficients in Eqs. (2-5) are listed in **Table 3** (see also [30] for more details). It is evident then that the closer a given location is to the nutrient source, the higher the levels of the three cues will be at this grid point. Glucose will be continuously taken up by cells to support their metabolism. Only the nutrient source, *p(150, 150)*, is replenished at each time step while all other grid points are not. In addition, cells take up both their own EGF and that secreted by adjoining cells in our model, because cancer cells act in both autocrine and paracrine manner in consuming EGF [40, 41]. (We note that for simplicity we treat both EGFR ligands, EGF and TGFα as being identical).

**Table 3**

Each cell encompasses a self-maintained molecular interaction network (shown in **Fig. 1**) and the simulation system records the molecular composite profile at every time step to determine the cell's phenotype for the next step. In between time steps, the chemical environment is being updated, including EGF and glucose concentration as well as oxygen tension (according to Eq. (5)). When the first cell reaches the nutrient source the simulation run is terminated.

**Cellular Phenotype Decision**

Four tumor cell phenotypes are considered in the model: proliferation, migration, quiescence and death. Cell death is triggered when the on site glucose concentration drops below 8 mM [42]. A cell turns quiescent when the on site glucose concentration is between 8 mM and 16 mM, when it does not meet conditions for migration or proliferation (see below), or when it cannot find an empty location to migrate or proliferate into.

The most important two phenotypic traits for spatio-temporal expansion, i.e. migration and proliferation, are decided by evaluating the dynamics of the following critical intracellular molecules. **(1)** *PLCγ* is known to be involved in directing cell movement in response to EGF [43-45]; *PLCγ* dynamics are accelerated during migration in cancer cells [46]. Therefore, in our model, the rate of change of PLCγ (ROC$_{PLC}$) decides if a cell proceeds to migration or not. That is, if ROC$_{PLC}$ exceeds a certain set threshold, T$_{PLC}$, the cell has the potential to migrate. **(2)** Similarly, the rate of change of *ERK* (ROC$_{ERK}$) decides if a cell proceeds with proliferation. ERK has been found experimentally to have a strong influence on cell proliferation [33, 47, 48], and transient activation of ERK with EGF leads to cell replication [49, 50]. If a cell decides to migrate or proliferate, it will search for an appropriate location to move to or for its offspring to reside in. Candidate locations are those grid points surrounding the cell. Implementing a cell surface receptor-mediated chemotactic evaluation, the most appropriate location is detected by using a 'search-precision' mechanism [27] according to:

$$T_{ij} = \psi \cdot L_{ij} + (1 - \psi) \cdot \varepsilon_{ij} \qquad (6)$$

where $T_{ij}$ represents the perceived attractiveness of location $p(i,j)$, $L_{ij}$ represents the result of an evaluation function for location $p(i,j)$ (see [27] for the definition of $L_{ij}$), and $\varepsilon \sim N(\mu, \sigma^2)$ is an error term following a normal distribution with mean $\mu$ and variance $\sigma^2$. $\psi \in [0,1]$ denotes the search-precision parameter that for a given run is held constant for all cells. Briefly, for a given cell at a certain location, when $\psi = 0$ the cell performs a pure random walk, whereas when $\psi = 1$ the cell always selects the location with the highest glucose concentration. Based on previous results [26], we set $\psi = 0.7$ because this value tends to lead to the highest average velocity of the tumor's spatial expansion.

It is worth noting that even if $ROC_{PLC}$ or $ROC_{ERK}$ exceed their corresponding thresholds, it does not necessarily have to lead to cell migration or proliferation. Rather, if nowhere else to go, the cell remains quiescent and continues to search for an empty location at the next time step.

**Figure 3**

Any cell in the process of changing its phenotype will fall into one of these four categories: (i) $ROC_{PLC} < T_{PLC}$ and $ROC_{ERK} < T_{ERK}$; (ii) $ROC_{PLC} > T_{PLC}$ and $ROC_{ERK} < T_{ERK}$; (iii) $ROC_{PLC} < T_{PLC}$ and $ROC_{ERK} > T_{ERK}$; and (iv) $ROC_{PLC} > T_{PLC}$ and $ROC_{ERK} > T_{ERK}$. **Figure 3** lists these conditions and their phenotypic consequences, respectively. Following the first three cell decisions is straightforward; first, if a cell experiences condition (i) no phenotypic change results as both $ROC_{PLC}$ and $ROC_{ERK}$ remain below their corresponding thresholds; however, if a cell faces condition (ii) the cell migrates because of $ROC_{PLC}$ exceeding its threshold while in the presence of

(iii) the cell proliferates due to $ROC_{ERK}$ exceeding its threshold. However, for (iv), and in the absence of any specific experimental data, i.e. for the case that *both* $ROC_{PLC}$ and $ROC_{ERK}$ exceed their corresponding thresholds, we explored two hypotheses: 'rule A' yielding migration advantage (i.e., the cell decides to migrate) whereas 'rule B' resulting in a proliferation advantage (i.e., the cell decides to proliferate). For simplicity, decision rules for the first three conditions are referred to '*general rules*', while rules A and B are referred to '*special rules*' hereafter. In the following section, we will describe the corresponding simulation results.

## Results

Our algorithm was implemented in C/C++. A total of 49 seed cells were initially set up in the center of the lattice, and these cells were arranged in a 7 x 7 square shape (i.e., M = 7 and N = 7, see **Fig. 2** for the configuration of the seed cells). We defined cell IDs from 0 to 48 (*left* to *right*, *bottom* to *top*). To investigate cell expansion dynamics, we monitored all cells and recorded their molecular profiles at every time step. We are particularly interested in the following four boundary cells: Cell No 0 (*bottom-left* corner, farthest from the source), Cell No 6 (*top-left* corner), Cell No 42 (*bottom-right* corner), and Cell No 48 (*top-right* corner, closest to the source). Through the distinct micro-environmental conditions they face, these corner cells exemplify the impact of location on single cell behavior, while they however still grasp the nature of the entire system. As described before, both rules A and B were tested for each different simulation condition (except the data reported in **Figures 6** and **7** which result from investigating rule A only).

**Multi-Cellular Dynamics**

**Figure 4** shows two simulation results for rules A and B, respectively. The simulations were conducted with a standard EGF concentration of 2.56 nM. Note that this concentration is derived from the literature [51, 52] and has been rescaled to fit our model as a benchmark starting point for further simulations. In the *upper panel* of **Fig. 4(a)** for rule A, tumor cells first display on site proliferation prior to exhibiting extensive migratory behavior towards the nutrient source. However, for rule B (*lower panel*), cells remain stationary proliferative throughout, thereby increasing the tumor radius yet without substantial mobility-driven spatial expansion. The run time for the latter case (rule B) was considerably longer than for rule A. Based on the criterion chosen for terminating the run, i.e. the first cell reaching the nutrient source, this result is somewhat expected since rule A favors migration whereas rule B promotes proliferation. This is further supported by analysis of the evolution of the various phenotypes and the change of [total] cell numbers (**Fig. 4(b)**). While both rules generate all three cell phenotypes (proliferation (*dark blue*), migration (*red*), and quiescence (*green*)), rule A (*left panel*) indeed appears to result in a cancer cell population that exhibits a larger migratory fraction than the one emerging through rule B (*right panel*) which, however, yields a larger portion of proliferative cells (*light blue*). It is thus not surprising that for rule B, the [total] cell population of the tumor system exceeds the one achieved through rule A by a factor of 10.

**Figure 4**

**Influence of Decision Rules on Phenotypic Changes**

To better understand the significance of each rule for the tumor system, we have investigated its influence on generating the intended phenotype. **Figure 5** shows the weight of rule A on migration **(a)**, and that of rule B on proliferation **(b)**. (The results are taken from the two simulation runs reported in **Fig. 4**). In **Fig. 5(a)**, migrations derive from two sources: (1) general rule, i.e. [$ROC_{PLC} > T_{PLC}$ and $ROC_{ERK} < T_{ERK}$] and (2) rule A; proliferations stem from one source only, i.e. if [$ROC_{PLC} < T_{PLC}$ and $ROC_{ERK} > T_{ERK}$]. Rule A plays a more dominant role in triggering migrations than the general rule does, yet does not contribute to increasing proliferations. Likewise, rule B has influence on proliferation only (**Fig. 5(b)**) and it contributes more to inducing proliferations than the corresponding general rule does too.

**Figure 5**

However, as documented in the linear least square fittings, the rate at which rule A causes an increase in migration exceeds by far the one by which rule B induces an increase in proliferation. This indicates that the influence of rule A on increasing migrations is *more* substantial than that of rule B on increasing proliferations. Being particularly interested in gaining insights into spatially aggressive tumors, we continue in the following with investigating the implications of rule A on microscopic and molecular level dynamics of the cancer system.

**Phase-Transition at Molecular Level**

To further investigate (for rule A) the relationship between EGF concentration and phenotypic changes we varied the extrinsic EGF concentration from the standard value of 2.65 x 1.0 nM to 2.65 x 50.0 nM by an incremental increase of 0.1 nM in each simulation. As a result of the model's underlying chemotactic search paradigm, expectedly a simulation under the condition of a higher extrinsic EGF concentration finished faster than that with a lower one. However, cells turn out not to exhibit completely homogeneous behavior.

Specifically, we focus on Cell No 48, the cell closest to the nutrient source, and report its corresponding molecular changes in **Fig. 6**. One can see that as the standard EGF concentration increases, the number of proliferations (*blue*) decreases gradually up to a *phase transition* between 2.65 x 31.1 and 2.65 x 31.2 nM. That is, if the standard EGF concentration is less than 2.65 x 31.1 nM, proliferation still occurs in this particular cell, but if the ligand concentration starts to exceed 2.65 x 31.2 nM, its proliferative trait entirely disappears. That is, in the presence of nutrient abundance, a very minor increase in extrinsic EGF can apparently abolish the expression of a phenotype. Even more intriguing, although the subcellular concentration change appears to be rather similar with regards to its patterns, on a closer look, the peak maxima of the rate changes for PLCγ and the turning point of the rate changes for ERK occur at an *earlier* time point for increasing EGF concentrations. This finding suggests that in the presence of excess ligand, the here implemented intracellular network switches to a *more efficient* signal processing mode. We note that for cell IDs 0, 6, and 42, no such phase transition emerged (data not shown) hence further supporting that this behavior is concentration dependent, and that geography, i.e. a cell's position relative to nutrient abundance, matters. Confirming the robustness of our finding for Cell No 48 we note that this cell continued to experience a phase transition when the coordinates of the center of the initial 49 cells was set randomly within a square region where *p(100,100)* is the lower left corner and *p(110,110)* is the upper right corner (5 runs, data not shown).

**Figure 6**

# Discussion & Future Works

While using mathematical models to investigate the behavior of signaling networks is hardly new, understanding a complex biosystem, such as a tumor, by focusing on the analysis of its molecular or cellular level separately or exclusively is insufficient, particularly if it excludes the interaction with the surrounding tissue. Recent analyses of signaling pathways in mammalian systems have revealed that highly connected sub-cellular networks generate signals in a context dependent manner [53]. That is, biological processes take place in heterogeneous and highly structured environments [54] and such extrinsic conditions alone can induce the transformation of cells independent of genetic mutations as has been shown for the case of melanoma [55]. Taken together, modeling of cancer systems requires the analysis and use of signaling pathways in a simulated cancer environment (context) across different spatial-temporal scales.

Our group has been focusing on the development of such multiscale models for studying highly malignant brain tumors [27, 29, 30, 56]. Here, on the basis of these previous works, we presented a 2D multiscale agent-based model to simulate NSCLC. Specifically, we monitored how, dependent on microenvironmental stimuli, molecular profiles dynamically change, and how they affect a single NSCLC cell's phenotype and, eventually, multicellular patterns.

Proceeding *top-down* in our analysis, we first evaluated the multicellular readout of molecular 'decision' rules A and B (versus general rules; **Fig. 3**). The patterns of a more stationary, concentrically growing cancer system (following rule B) are quite different from the rapid, chemotactically-guided, spatial expansion that can be seen in the tumor regulated by rule A (**Fig. 4(a)**). Not surprisingly, the latter also operates with many more migratory albeit overall less [total] cells (**Fig. 4(b)**). Furthermore, examining in more detail the influence of the two distinct rules on their respective phenotypic yield, we found that the impact of rule A on increasing cell migration is more substantial than rule B's influence on furthering proliferation (**Fig. 5**). This finding suggests that the migratory rule A can operate the cancer system through incrementally smaller changes (while the simulation system is more robust for rule B). Such *sensitivity* to migratory cues corresponds well with experimental data on the response of human breast cancer cells, which showed that a spatially successful expansive system reacts rather quickly to even miniscule changes in chemotactic directionality [57, 58].

Continuing therefore with rule A, our effort was then geared to gain insights into tumor expansion dynamics not only with regards to extrinsic stimuli but also to cell geography, i.e. a cell's location relative to the replenished nutrient source. Most interestingly, we found a *phase transition* in the cancer cell closest to the nutrient source (i.e. Cell No 48, while none of the other three corner cells showed similar behavior). Specifically, for a tumor cell at this location, i.e., facing nutrient abundance, proliferation is completely *abolished* once the extrinsic EGF concentration exceeds a certain level. While this at first may seems rather unexpected, this finding however only confirms the experimentally sound notion that EGF stimulates the spatial expansion of a cancer system [5-8]. Moreover, with increasing EGF concentrations, the maxima of $ROC_{PLC}$ (**Fig. 6**) gradually occur *earlier* which seems to indicate that, under these conditions, the downstream signal is processed *faster*. Interestingly, such a 'no proliferation, just migration' behavior in the presence of chemo-attractant has indeed already been reported in several *in vitro* studies using a variety of cancer cell lines [59, 60] as well as in non-cancerous human cells [61]. (While admittedly, for the reasons stated, rule B did not receive similar attention in our analysis), we nonetheless argue that, on the basis of our results and the experimental reports they seem to correspond with, rule A and thus a *migratory* decision prompted by a [$ROC_{PLC} > T_{PLC}$ and $ROC_{ERK} > T_{ERK}$] condition is a reasonable outcome for the signaling process taking place in NSCLC also *in vitro* and *in vivo*.

However, moving the model closer to reality will require a multitude of adjustments, one of which is its ability to account for up- or down-regulation in key molecules as a result of tumorigenesis. As a first step, and since experimental data on over-expression of EGFR in a variety of cancer types, including NSCLC, are ample [62-65] we have begun to simulate the impact of an increasing number of receptors on the cancer system (**Fig. 7**; simulations conducted with an EGFR concentration of 800 nM (per system)). Comparing this preliminary data with those reported in **Fig. 6** (simulations conducted with an EGFR concentration of 80 nM (per system)), we find that an EGFR-overexpressing NSCLC tumor seems to operate with even *more* migration and does so *earlier* on. The result is a spatially even more aggressive cancer system, which seems to correspond well with the aforementioned experimental studies. And, intriguingly, while the phase transition itself is preserved, it however occurs already at a smaller EGF concentration, hence indicating that the increase in receptor density leads to an *amplification* of the downstream signal, which again corresponds well with experimental results in examining signaling activities generated by different EGFR family members [66]. Taken together, while preliminary, this finding demonstrates applicability and confirms flexibility of this multiscale platform, hence warrants its further expansion.

**Figure 7**

There are a number of research tracks that can and should be pursued in future works. First, it will be intriguing to see if, in the presence of a *non*-replenished nutrient source, the proliferative phenotype eventually can be recovered once extrinsic ligand concentrations fall beyond the phase-transition threshold. More generally, while most of the pathway's parameters, including rate constants and initial component concentrations were obtained from the experimental literature, this data naturally originated from a variety of often stationary experimental settings and different cell types. It therefore represents a less desirable and reliable input than time series data that come from *one* experimental setting only. Also, some parameters had to be estimated, much like in other well-established pathway models [11, 12]. Taken together, future works will have to include not only proper experimental verification of the estimated parameters and evaluation of the simulation results but also, on the *in silico* side, techniques such as sensitivity analysis to help determine the effects of parameter uncertainties on model outcome [67] and to identify control points for experiment design [68]. While a pathway model cannot be a biological representation in every detail [38] we plan on adding, in incremental steps, other pathways of relevance for NSCLC such as e.g. PI3K/PTEN/AKT [69]. Moreover, simulating a more heterogeneous biochemical environment and implementing both cell-cell and cell-matrix interactions [70] are planned steps at the cellular level that should help representing the cancer system of interest in more detail.

Regardless, we believe that the current model already provides useful insights into NSCLC from a systematic view in terms of quantitatively understanding the relationship between extrinsic chemotactic stimuli, the underlying properties of signaling networks, and the cellular biological responses they trigger. Our results yield several experimentally testable hypotheses and thus further support the use of multiscale models in interdisciplinary cancer research. To our knowledge, this presents the first multiscale computational model of Non-Small Cell Lung Cancer and is thus potentially a significant first step towards realizing a fully validated *in silico* model for this devastating disease.

## List of abbreviations used

EGF = epidermal growth factor; EGFR = EGF receptor; ERK = extracellular signal-regulated kinase; MAPK = mitogen-activated protein kinase; MEK = MAPK kinase; NSCLC = non-small cell lung cancer; PLCγ = phospholipase Cγ; PKC = protein kinase C; TGFα = transforming growth factor α.

# Competing interests

The authors declare that they have no competing interests.

# Authors' contributions

ZW developed the NSCLC model (algorithm and code), analyzed its simulation results and drafted the manuscript. LZ assisted in developing the algorithm, while JS supported data analysis and preparation of manuscript. TSD developed the model's underlying concept, led its design, development and analysis, and finalized the manuscript. All authors read and approved the final manuscript.

# Acknowledgements

This work has been supported in part by NIH grant CA 113004 (The Center for the Development of a Virtual Tumor, CViT at http://www.cvit.org) and by the Harvard-MIT (HST) Athinoula A. Martinos Center for Biomedical Imaging and the Department of Radiology at Massachusetts General Hospital. We would like to acknowledge helpful discussions with Drs. Raju Kucherlapati, Victoria Joshi and David Sarracino at Harvard-Partners Center for Genetics and Genomics (HPCGG).

# Figures

**Figure 1  - Kinetic model of the NSCLC-specific EGFR signaling pathway**

The *arrows* represent the reactions specified in **Tables 1** and **2**.

**Figure 2 - Two-dimensional virtual micro-environment**

Depicted are the 200 x 200 lattice (*left*) with the position of the nutrient source, and the seed cells with assignment of the corner cell IDs (0, 6, 42, and 48).

**Figure 3  - Cell phenotypic decision algorithm**

See text for more details.

**Figure 4  - Multicellular tumor expansion dynamics**

(**a**) Shows the multicellular patterns that emerge through rule A (*upper panel*) and rule B (*lower panel*), respectively. (**b**) Describes the numeric evolution (*y-axis*) of each cell phenotype as well as of the [total] cell population (*light blue*) over time (*x-axis*) for rule A (*left panel*) and rule B (*right panel*), respectively. Note: proliferative tumor cells are labeled in *dark blue*, migratory cells in *red*, quiescent cells in *green* and dead cells in *grey*.

**Figure 5  - Weight of decision rules on changing cell phenotypes**

Influence on changing cell migration (*left panel*) and proliferation (*right panel*) when following the corresponding rule (see **Fig. 3**). The *dashed red* line indicates rule A-mediated migrations in (**a**), while the *dashed blue* line denotes rule B-mediated proliferations in (**b**). Fitting curves in *solid black* are calculated using a standard linear least squares method. Slopes of the fitting curves are 1.40 cells/step in (**a**) and 0.03 cells/step in (**b**), respectively. Note: The drop of the dashed red line in the *left panel* of (**a**) is caused by the termination of the simulation when a cell reached the source (in this case, no further computation on remaining cells will be performed).

**Figure 6  - Changes at the molecular level for Cell No 48 with an increasing extrinsic EGF concentration (rule A)**

Four simulation runs are depicted where (from *left to right*) the EGF concentration increases from 2.65 x 1.0 to 2.65 x 31.1, 2.65 x 31.2, and finally, to 2.65 x 50.0 nM. (From *top* to *bottom*) plotted are the absolute change of PLCγ, rate of change of PLCγ, and rate of change of ERK. Note that the number of proliferations is decreasing gradually and finally disappears at a phase transition between the EGF concentrations of 2.65 x 31.1 and 2.65 x 31.2 nM. (For phenotype labeling see **Fig. 4**).

**Figure 7  - Changes at the molecular level for Cell No 48 with an increasing extrinsic EGF concentration (rule A), at an EGFR concentration of 800 nM**

Three simulation runs are depicted where (from *left to right*) the EGF concentration increases from 2.65 x 1.0 to 2.65 x 5.9 and 2.65 x 6.0 nM. (From *top* to *bottom*) plotted are the absolute change of PLCγ, rate of change of PLCγ, and rate of change of ERK. Note that a phase transition emerges again between the EGF concentrations of 2.65 x 5.9 and 2.65 x 6.0 nM, hence at a *lower* concentration compared to the one depicted in **Fig. 6** (EGFR concentration of 80 nM) . In the two simulations around the phase transition, the maximum rates of change for both PLCγ and ERK (i.e., ROC$_{PLC}$ and ROC$_{ERK}$ at 2.65 x 5.9 and at 2.65 x  6.0 nM) are *lower* compared with those in **Fig. 6** (i.e., ROC$_{PLC}$ and ROC$_{ERK}$ at 2.65 x 31.1 and at 2.65 x  31.2 nM). (For phenotype labeling see **Fig. 4**).

## FIGURE 1.

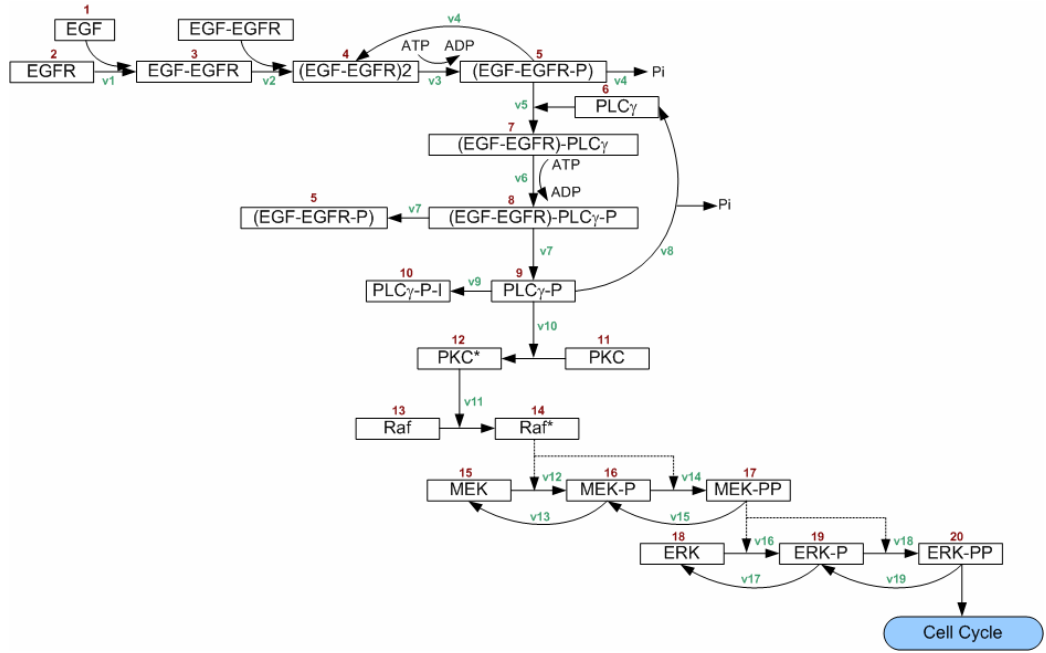

**FIGURE 2.**

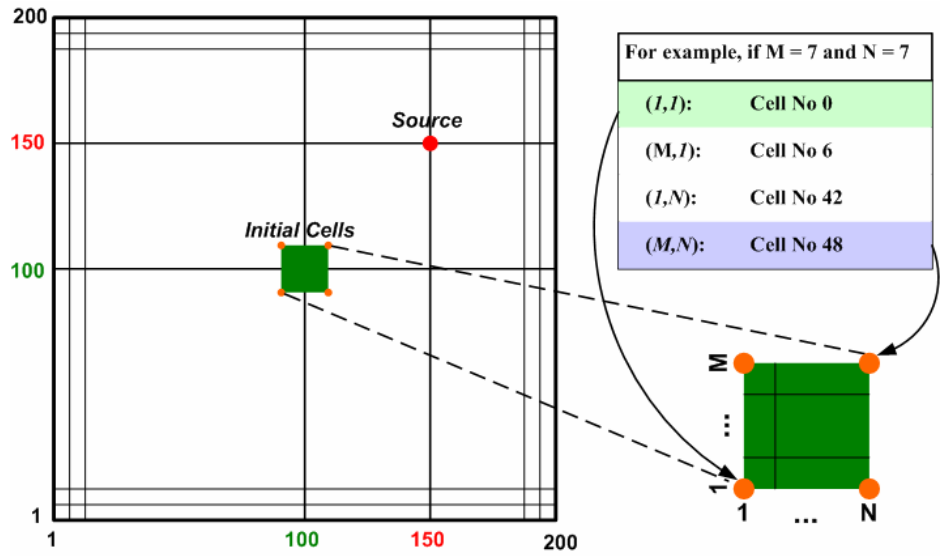

**FIGURE 3.**

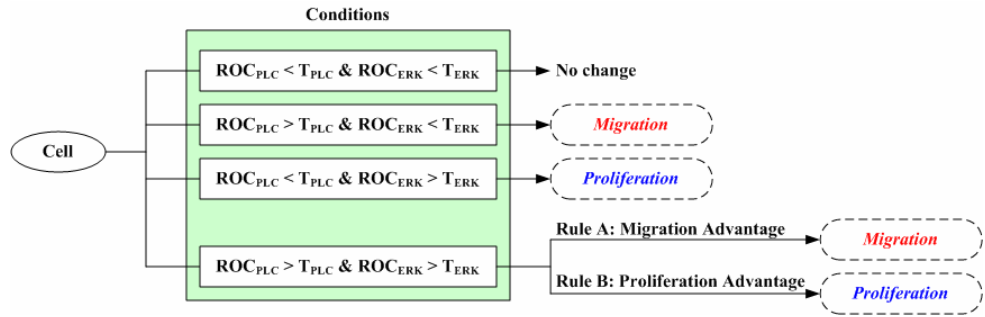

**FIGURE 4.**

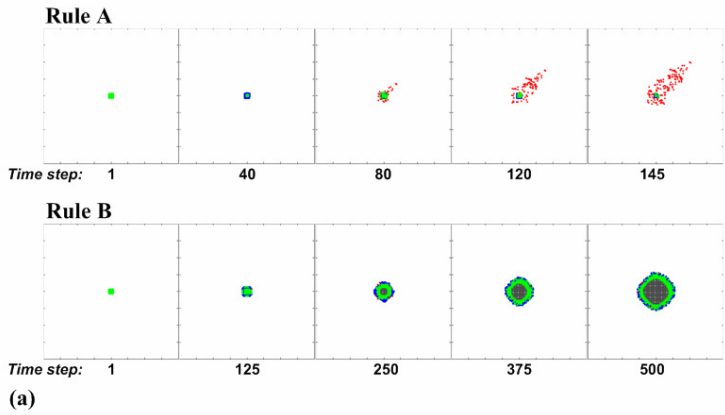

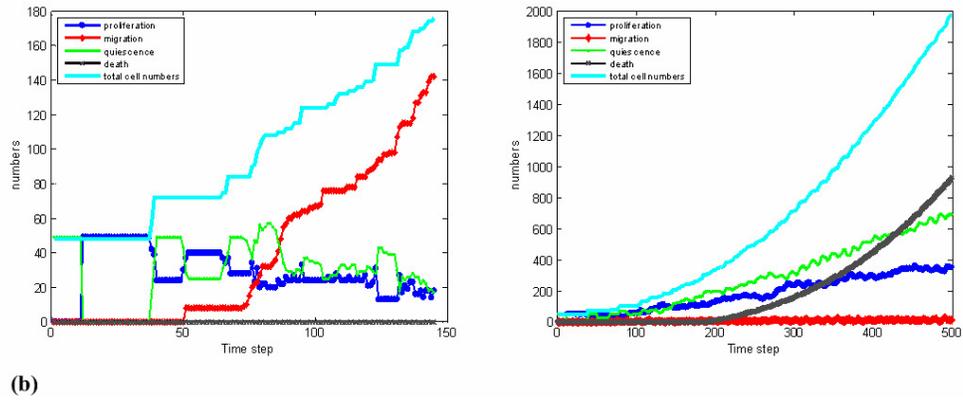

**FIGURE 5.**

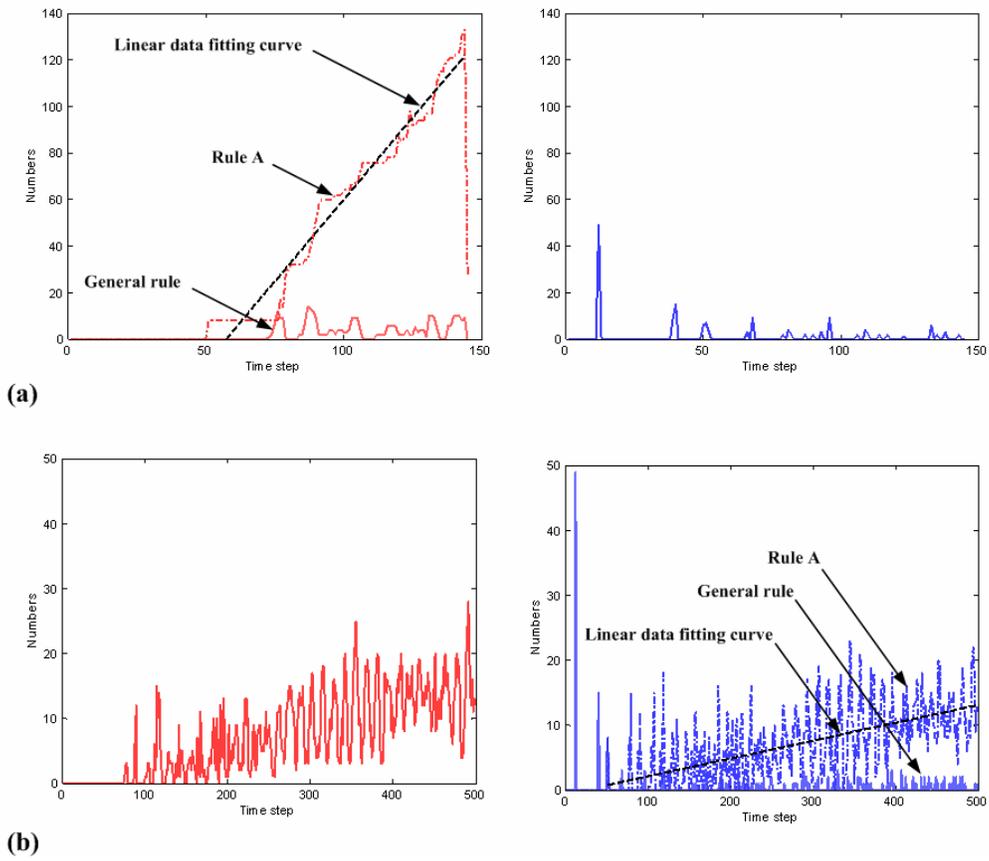

**FIGURE 6.**

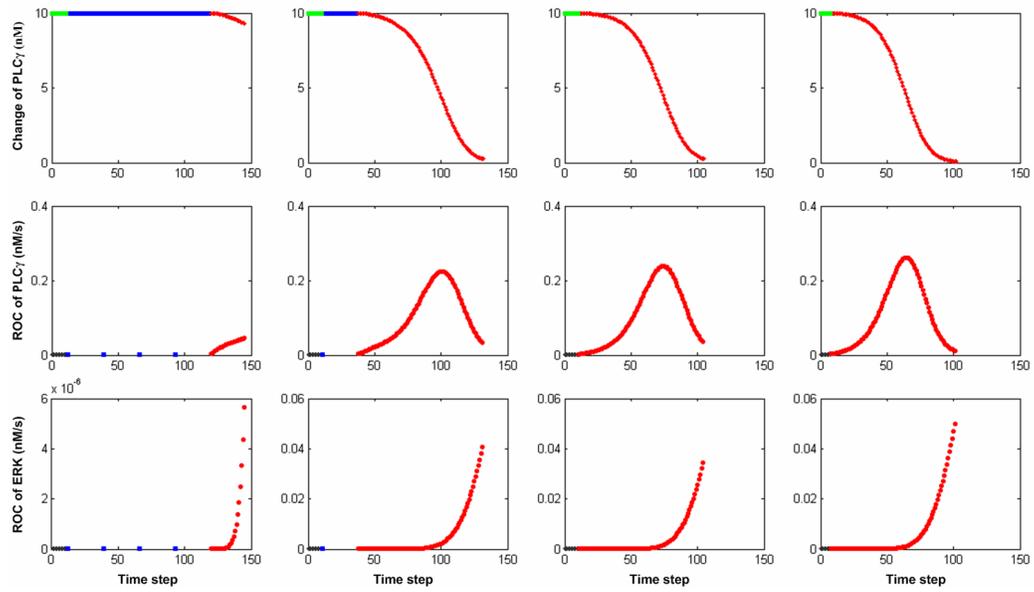

**FIGURE 7.**

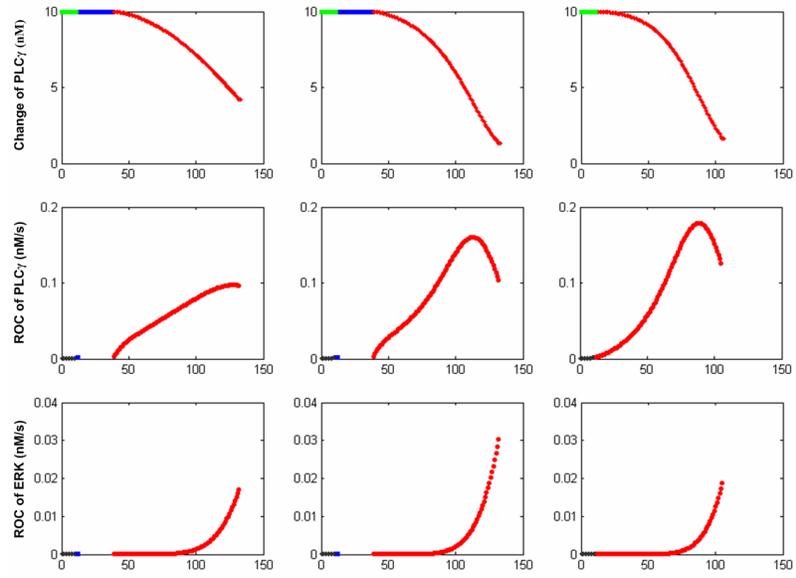

# Tables

### Table 1 - Kinetic equations and initial concentrations

See **Table 2** for references.

| Reactant | Molecular variable | Initial concentration [nM] | ODE |
|---|---|---|---|
| $X_1$ | EGF | to be varied | $d(X_1)/dt = -v_1$ |
| $X_2$ | EGFR | 80 | $d(X_2)/dt = -v_1$ |
| $X_3$ | EGF-EGFR | 0 | $d(X_3)/dt = v_1 - 2v_2$ |
| $X_4$ | (EGF-EGFR)2 | 0 | $d(X_4)/dt = v_2 + v_4 - v_3$ |
| $X_5$ | EGF-EGFR-P | 0 | $d(X_5)/dt = v_3 + v_7 - v_4 - v_5$ |
| $X_6$ | PLC$\gamma$ | 10 | $d(X_6)/dt = v_8 - v_5$ |
| $X_7$ | EGF-EGFR-PLC$\gamma$ | 0 | $d(X_7)/dt = v_5 - v_6$ |
| $X_8$ | EGF-EGFR-PLC$\gamma$-P | 0 | $d(X_8)/dt = v_6 - v_7$ |
| $X_9$ | PLC$\gamma$-P | 0 | $d(X_9)/dt = v_7 - v_8 - v_9 - v_{10}$ |
| $X_{10}$ | PLC$\gamma$-P-I | 0 | $d(X_{10})/dt = v_9$ |
| $X_{11}$ | PKC | 10 | $d(X_{11})/dt = -v_{10}$ |
| $X_{12}$ | PKC* | 0 | $d(X_{12})/dt = v_{10} - v_{11}$ |
| $X_{13}$ | Raf | 100 | $d(X_{13})/dt = -v_{11}$ |
| $X_{14}$ | Raf* | 0 | $d(X_{14})/dt = v_{11} - v_{12} - v_{14}$ |
| $X_{15}$ | MEK | 120 | $d(X_{15})/dt = v_{13} - v_{12}$ |
| $X_{16}$ | MEK-P | 0 | $d(X_{16})/dt = v_{12} + v_{15} - v_{13} - v_{14}$ |
| $X_{17}$ | MEK-PP | 0 | $d(X_{17})/dt = v_{14} - v_{15} - v_{16} - v_{18}$ |
| $X_{18}$ | ERK | 100 | $d(X_{18})/dt = v_{17} - v_{16}$ |
| $X_{19}$ | ERK-P | 0 | $d(X_{19})/dt = v_{16} + v_{19} - v_{17} - v_{18}$ |
| $X_{20}$ | ERK-PP | 0 | $d(X_{20})/dt = v_{18} - v_{19}$ |

- PKC* and Raf * indicate the activated form of PKC and Raf, respectively.

**Table 2 - Kinetic parameters**

Concentrations and the Michaelis-Menten constants ($K_4$, $K_8$, and $K_{11}$–$K_{19}$) are given in [nM]. First- and second-order rate constants are given in [$s^{-1}$] and [$nM^{-1} \cdot s^{-1}$], respectively. $V_4$, $V_8$, and $V_{11}$–$V_{19}$ are expressed in [$nM \cdot s^{-1}$].

| Reaction number | Equation | Kinetic parameter | | Reference |
|---|---|---|---|---|
| $v_1$ | $k_1 \cdot X_1 \cdot X_2 - k_{-1} \cdot X_3$ | $k_1{=}0.003$ | $k_{-1}{=}0.06$ | [11] |
| $v_2$ | $k_2 \cdot X_3 \cdot X_3 - k_{-2} \cdot X_4$ | $k_2{=}0.01$ | $k_{-2}{=}0.1$ | [11] |
| $v_3$ | $k_3 \cdot X_4 - k_{-3} \cdot X_5$ | $k_3{=}1$ | $k_{-3}{=}0.01$ | [11] |
| $v_4$ | $V_4 \cdot X_5 / (K_4 + X_5)$ | $V_4{=}450$ | $K_4{=}50$ | [11] |
| $v_5$ | $k_5 \cdot X_5 \cdot X_6 - k_{-5} \cdot X_7$ | $k_5{=}0.06$ | $k_{-5}{=}0.2$ | [11] |
| $v_6$ | $k_6 \cdot X_7 - k_{-6} \cdot X_8$ | $k_6{=}1$ | $k_{-6}{=}0.05$ | [11] |
| $v_7$ | $k_7 \cdot X_8 - k_{-7} \cdot X_5 \cdot X_9$ | $k_7{=}0.3$ | $k_{-7}{=}0.006$ | [11] |
| $v_8$ | $V_8 \cdot X_9 / (K_8 + X_9)$ | $V_8{=}1$ | $K_8{=}100$ | [11] |
| $v_9$ | $k_9 \cdot X_9 - k_{-9} \cdot X_{10}$ | $k_9{=}1$ | $k_{-9}{=}0.03$ | [11] |
| $v_{10}$ | $k_{10} \cdot X_9 \cdot X_{11} - k_{-10} \cdot X_{12}$ | $k_{10}{=}0.214$ | $k_{-10}{=}5.25$ | Estimate |
| $v_{11}$ | $V_{11} \cdot X_{12} \cdot X_{13} / (K_{11} + X_{13})$ | $V_{11}{=}4$ | $K_{11}{=}64$ | [39] |
| $v_{12}$ | $V_{12} \cdot X_{14} \cdot X_{15} / [K_{12} \cdot (1 + X_{16} / K_{14}) + X_{15}]$ | $V_{12}{=}3.5$ | $K_{12}{=}317$ | [14] |
| $v_{13}$ | $V_{13} \cdot X_{16} / [K_{13} \cdot (1 + X_{17} / K_{15}) + X_{16}]$ | $V_{13}{=}0.058$ | $K_{13}{=}2200$ | [12] |
| $v_{14}$ | $V_{14} \cdot X_{14} \cdot X_{16} / [K_{14} \cdot (1 + X_{15} / K_{12}) + X_{16}]$ | $V_{14}{=}2.9$ | $K_{14}{=}317$ | [12] |
| $v_{15}$ | $V_{15} \cdot X_{17} / [K_{15} \cdot (1 + X_{16} / K_{13}) + X_{17}]$ | $V_{15}{=}0.058$ | $K_{15}{=}60$ | [12] |
| $v_{16}$ | $V_{16} \cdot X_{17} \cdot X_{18} / [K_{16} \cdot (1 + X_{19} / K_{18}) + X_{18}]$ | $V_{16}{=}9.5$ | $K_{16}{=}1.46 \times 10^5$ | [12] |
| $v_{17}$ | $V_{17} \cdot X_{19} / [K_{17} \cdot (1 + X_{20} / K_{19}) + X_{19}]$ | $V_{17}{=}0.3$ | $K_{17}{=}160$ | [12] |
| $v_{18}$ | $V_{18} \cdot X_{17} \cdot X_{19} / [K_{18} \cdot (1 + X_{18} / K_{16}) + X_{19}]$ | $V_{18}{=}16$ | $K_{18}{=}1.46 \times 10^5$ | [12] |
| $v_{19}$ | $V_{19} \cdot X_{20} / [K_{19} \cdot (1 + X_{19} / K_{17}) + X_{20}]$ | $V_{19}{=}0.27$ | $K_{19}{=}60$ | [12] |

**Table 3 - Coefficients of distribution and diffusion of EGF, glucose and oxygen tension**

Values are taken from the literature [71, 72].

| Coefficient | Value | Units | Description |
|---|---|---|---|
| $T_m$ | 2.56 | nM | Maximum concentration of EGF |
| $G_a$ | 17.0 | mM | Normal concentration of glucose |
| $G_m$ | 57.0 | mM | Maximum concentration of glucose |
| $O_a$ | 0.0017 | DC | Normal concentration of oxygen |
| $O_m$ | 0.0025 | DC | Maximum concentration of oxygen |
| $D_{EGF}$ | $6.7 \times 10^{-11}$ | $m^2 \cdot s^{-1}$ | Diffusion coefficient of EGF |
| $D_{Glucose}$ | $5.18 \times 10^{-11}$ | $m^2 \cdot s^{-1}$ | Diffusion coefficient of glucose |
| $D_{Oxygen}$ | $8.0 \times 10^{-9}$ | $m^2 \cdot s^{-1}$ | Diffusion coefficient of oxygen |